\documentclass[twocolumn,superscriptaddress,showpacs]{revtex4}
\usepackage{amsmath}
\usepackage{amsfonts}
\usepackage{amssymb}
\usepackage{graphicx}
\usepackage{epstopdf}
\usepackage{times}
\usepackage{subfigure}
\usepackage{verbatim}
\usepackage{bm}

\newcommand{\dissip}{\epsilon}

\begin{document}

\title{Lagrangian view of time irreversibility of fluid turbulence}

\author{Haitao Xu}
\email{hxu@tsinghua.edu.cn}
\affiliation{Center for Combustion Energy and Department of Thermal Engineering, Tsinghua University, Beijing 100084, China}
\affiliation{Max Planck Institute for Dynamics \& Self-Organization (MPIDS), D-37077 G\"ottingen, Germany}

\author{Alain Pumir}
\affiliation{Ecole Normale Sup\'erieure de Lyon, F-69007 Lyon, France}
\affiliation{Max Planck Institute for Dynamics \& Self-Organization (MPIDS), D-37077 G\"ottingen, Germany}

\author{Eberhard Bodenschatz}
\affiliation{Max Planck Institute for Dynamics \& Self-Organization (MPIDS), D-37077 G\"ottingen, Germany}
\affiliation{Institute for Nonlinear Dynamics, University of G\"ottingen, D-37077 G\"ottingen, Germany}
\affiliation{Laboratory of Atomic and Solid State Physics and Sibley School of Mechanical and Aerospace Engineering, Cornell University, Ithaca, NY 14853, USA}

\begin{abstract}
A turbulent flow is maintained by an external supply of kinetic energy, which is 
eventually dissipated into heat at steep velocity gradients. The scale at which energy is supplied greatly differs
from the scale at which energy is dissipated, the more so
as the turbulent intensity (the Reynolds number) is larger. 
The resulting energy flux
over the range of scales, intermediate between energy injection and dissipation, 
acts as a source of time irreversibility.
As it is now possible to follow accurately fluid particles in a turbulent flow field, 
both from laboratory experiments and from numerical simulations, a natural question 
arises: how do we detect time irreversibility from these Lagrangian data? 
Here we discuss recent results concerning this problem. 
For Lagrangian statistics involving more than one fluid particle, the distance 
between fluid particles introduces an intrinsic length scale into the 
problem. 
The evolution of quantities dependent on the relative motion between these 
fluid particles, including the kinetic energy in the relative motion, 
or the configuration of an initially isotropic 
structure can be related to the equal-time 
correlation functions of the velocity field, and is therefore sensitive to the energy flux through 
scales, hence to the irreversibility of the flow. 
In contrast, for single-particle Lagrangian statistics, 
the most often studied velocity structure functions
cannot distinguish the ``arrow of time''.
Recent observations from experimental and numerical 
simulation data, however, show that the change of kinetic energy 
following the particle motion, is sensitive to time-reversal. 
We end the survey with a brief discussion of the implication of this line of work.
\end{abstract}


\maketitle

\section{Introduction}

Flowing fluids are ubiquitous in many natural and industrial situations.
The running of water out of the faucet in our kitchen or the 
intimidating roar of a destructive hurricane provide 
examples involving the two most abundant fluids on earth: water and air.
From elementary physical principles, the description of 
fluid motions is based on the competition between the inertia of fluid 
particles, and the diffusion of momentum by viscosity. As a result, the 
physical properties of the flow are characterized by the 
dimensionless Reynolds number $Re = U L / \nu$, where $L$ and $U$ are the 
typical length and velocity scales of the flow and $\nu$ is the kinematic viscosity of 
the fluid. The Reynolds number can be regarded as the ratio of the viscous 
time scale $L^2/\nu$ and the flow time scale $L/U$ and therefore measures the 
relative importance of the inertial effect, which tends to drive the flow to 
become unstable, and the damping by the viscosity. For a flow at small 
Reynolds number, viscous diffusion is fast, so viscosity
damps out flow disturbances. This situation is
referred to as ``laminar''.
In such flows, energy dissipation transfers mechanical energy
into heat, which can be readily seen from the few available exact solutions of 
the Navier-Stokes equations~\cite{Landau_FM}.
On the other hand, when the Reynolds number is large, the inertial effects 
dominate and the flow appears to be much more irregular, even 
in the absence of any externally imposed time-dependence. 
Such flows are called ``turbulent''. 
In three dimensional situations, turbulent flows are ``rough'', 
in the sense that they develop strong variations of the velocity field over
very small scales, or equivalently, very large velocity gradients. 
In these regions, the viscosity is important. 
The irregular nature of turbulent flows leads to a
fast and seemingly erratic motion of small particles transported by turbulence.
The work reviewed here shows that the
fundamental properties of the flow, such as the irreversibility induced by
the energy dissipation, manifest themselves in the motion of small particles. 

Most macroscopic flows in nature and technology are turbulent. 
This is a consequence of the very small values of the viscosities of
the most common fluids, such as air and water, which leads to large
Reynolds numbers, even
at modest length and velocity scales. For example, an adult walking at a
moderate pace creates an air flow with a Reynolds number of approximately 
$5 \times 10^4$ around him/her, and the flow of tap water in our kitchen can 
easily reach a Reynolds number of $10^4$, both of which are well in the 
turbulent flow regime.

Because of their rapid erratic motion, turbulent flows strongly enhance mixing.
This we know well from stirring water in order to dissolve sugar added in it. 
The same principle we apply when we rapidly mix fuel and air in combustion 
engines. This of course comes at the expense of energy: 
The intense, erratic turbulent flow needs to be maintained by external driving or pumping. 
Sometimes one would like to avoid turbulence. 
Examples can be found in fluid drag on trains, cars, ships and airplanes, or in pipe flows,
where a much larger pressure drop
is required to pump a turbulent flow through a pipe than a laminar flow 
at the same flow rate. 
Thus a better understanding of turbulence and turbulent 
flows could allow us, on the one hand, to mix fluids more efficiently and, on the other hand, 
to reduce the drag in technical applications.

From the point of view of fundamental physics,
turbulence is an emblematic example of non-equilibrium systems,
whose description is notoriously challenging. 
In particular, novel concepts to master the underlying complexity of turbulent
flows are yet to be discovered.
These concepts could then also provide deeper insights into 
other non-equilibrium problems.
One of the unifying concepts to address these problems is the breaking of
detailed balance. Irreversibility in turbulence implies that the transition
probabilities from a state A to a state B, and from the state B to the state A
are {\it not} equal, contrary to what happens in equilibrium systems in 
statistical mechanics~\cite{Landau_SP,Derrida:2007,CJP10}.

Like other out-of-equilibrium systems, turbulent fluid flows are irreversible. 
The kinetic energy of the flow is always dissipated and a constant supply of kinetic energy is necessary to maintain a turbulent flow. 
Whereas it is a simple matter to understand dissipation (hence 
irreversibility) in laminar flows, it is much more challenging to identify
the features of the turbulent flow motions that reveal irreversibility. 
In turbulent flows, the nonlinearity plays a crucial role, and leads to 
chaotic motion, with subtle statistical properties.
A specific property of turbulent flows is that the scales at which
energy is supplied, either from external forcing or from flow 
instabilities, are vastly different from the scales at which the energy is 
dissipated. In three-dimensional (3D) flows, the energy injection is at large 
scales while the viscous dissipation dominates at 
small-scales~\cite{TennekesLumley:1972,frisch:1995,pope:2000}. In two-dimensional (2D) flows, the energy is supplied at small scales and then dissipated by friction at large scales~\cite{K67,BE12}. The scale-separation between the energy injection and energy dissipation implies that there is, on average, a flux of kinetic energy through spatial scales between the energy injection scale and the energy dissipation scale, i.e., in the so-called inertial range, and the direction of this average energy flux cannot be reversed. This is the underlying reason why turbulent flows are irreversible. There are fundamental differences between fluctuations in steady-state turbulence and fluctuations in equilibrium systems~\cite{RS78}.

Remarkably, the celebrated Kolmogorov's 4/5-law shows that in statistically stationary turbulent flows, 
the energy flux can be measured from a single snapshot of the velocity field~\cite{K41a,K41c}, 
without any explicit reference to its temporal evolution. 
The 4/5-law can be generalized to both 2D and 3D cases as (see e.g.~Ref.~\cite{FGV01})
\begin{equation}
\left\langle \left[ \Big( \mathbf{u}(\mathbf{x} + \mathbf{r}, t) - \mathbf{u}(\mathbf{x}, t) \Big)\cdot \frac{\mathbf{r}}{r} \right]^3 \right\rangle = - \frac{12}{d (d+2)} \dissip r ,
\label{eq:45law}
\end{equation}
where
$\mathbf{u}(\mathbf{x},t)$ is the velocity field, $\mathbf{r}$ is the 
separation vector whose magnitude is $r$,
$d$ is the spatial dimension, and $\dissip$ is the turbulent energy flux, defined as a signed quantity, with the convention that $\dissip > 0$ when energy is transferred towards small scales and $\dissip < 0$ for energy flux toward large scales.
The symbol $\langle \cdot \rangle$ 
refers to an \emph{ensemble average}, obtained by averaging over all 
possible flow realizations.  For statistically homogeneous flows, 
it is the same as averaging over time $t$ (under the assumption of ergodicity), or 
over space $\mathbf{x}$. 
Equation~\eqref{eq:45law} establishes that the third moment
of the longitudinal velocity increments, 
$\delta_r u \equiv \Big( \mathbf{u}(\mathbf{x} + \mathbf{r}, t) - \mathbf{u}(\mathbf{x}, t) \Big)\cdot (\mathbf{r}/r)$, 
differs from zero, which is a consequence of the existence of
a flux of energy through scales. 
This equation is valid when the distance $r$ between the two measured velocities is in the inertial range, i.e., when $\ell_s \ll r \ll \ell_L$, where $\ell_s$ and $\ell_L$ are the smallest and the largest length scales of the flow whose physical meanings depend on the spatial dimension of the flow. For 3D turbulence, $\ell_s = \eta$, the well-known Kolmogorov scale given by the balance between the viscous dissipation and the energy flux $\dissip$ 
(remember that $\dissip$ can be either positive or negative, depending
on the nature of the transfer of energy),
and $\ell_L$ is the forcing scale. In the 3D turbulence case, $\dissip > 0$ and the flux is from the large to the small scales and it is called \emph{direct energy cascade}. 
On the other hand, for 2D turbulence, the energy flux is negative: $\dissip < 0$ and it is called \emph{inverse energy cascade} because the kinetic energy is from the forcing scale $\ell_f$, which is the small scale $\ell_s$, to the large scale $\ell_L$ that is determined by the balance between large-scale friction and the energy flux.

We stress that in Eq.~\eqref{eq:45law}, both $\mathbf{u}(\mathbf{x}+\mathbf{r})$ and 
$\mathbf{u}(\mathbf{x})$ are measured at the \emph{same} time $t$, 
i.e., in principle using only a single ``snapshot'' of the velocity field.
This property leads to the following paradoxical situation.
Consider a sequence of velocity fields, which correspond to a
solution of the Navier-Stokes equation. According to Eq.~\ref{eq:45law}, the 
precise order of the snapshots is immaterial in the determination of the rate
of energy dissipation. In particular, simply reversing the order of the
sequence, i.e., changing $ t \rightarrow -t$ does not seem to affect
the determination of the energy dissipation rate, which is obtained
from the individual velocity field (Eulerian statistics) alone.
Thus, we can measure the energy flux correctly but cannot detect the time-irreversibility, 
although the energy flux is the cause of the irreversibility. 

To understand this paradoxical situation, we investigate here
the motion of fluid particles moving with the flow, i.e., Lagrangian statistics. In the last two decades, significant advances in measurement techniques and computing technologies made it possible to obtain well-resolved Lagrangian statistics at high Reynolds numbers in both experiments~\cite{MOA99,laPorta01,MMMP01,LTK05,BOX+06,VMVP08,XPB11} and numerical simulations~\cite{YP89,PSC00,YB2004,BBC+05b,HYS:2011}. 
Unexpectedly, as explained in Section~\ref{sec:single}, the 
Lagrangian structure functions turn out to be completely  insensitive to the 
fundamental irreversibility of the flow~\cite{FXP+13}, thus calling for new
concepts and ideas.
In the following, we review briefly these recent results, with the focus on the relation to the irreversibility of turbulence.

\section{Multi-particle statistics: The role of energy flux}

We start with multi-particle Lagrangian statistics, i.e., 
by following more than one fluid particle in a turbulent flow. 
To this end, we consider elementary sets of particles, and 
study the flow perceived by the particles, and its dependence
on the characteristic distance between the particles.
We show that when the inter-particle separations 
are in the inertial range, the energy flux through scales results
in a measurable difference between the statistical properties of the 
motion forward and backward in time.

\subsection{Relative dispersion and energy considerations}

In a turbulent flow, two fluid particles are, on average, moving away from each other.
This can be quantified by measuring the mean squared stance between the two particles, 
$\langle R^2(t) \rangle \equiv \langle \vert \mathbf{X}_1(t) - \mathbf{X}_2 (t) \vert^2 \rangle_{R_0}$ 
where $\mathbf{X}_1(t)$ and $\mathbf{X}_2(t)$ are the positions of particles 
1 and 2 at time $t$. 
The subscript $R_0$ in the definition of the mean squared distance between
two particles, $\langle R^2(t) \rangle$, refers to the imposed condition
that at $t=0$ the initial distance $R(0) = | \mathbf{X}_1 (0) - \mathbf{X}_2(0) | $
is equal to $R_0$.

How $\langle R^2(t) \rangle$ changes with time quantifies
the relative dispersion of a particle pair. The problem
has been extensively studied since the pioneering work by 
Richardson~\cite{Richardson1926}, who observed that in turbulent flows,
in a meteorological context, the mean squared 
distance between two particles grows with time as $t^3$, i.e., 
$\langle R^2(t) \rangle \propto t^3$. Paradoxically, 
although Richardson $t^3$ law is easy to justify theoretically,
it has proven very difficult to observe this regime
of turbulent dispersion
in any other well-controlled laboratory 
flows, or in direct numerical 
simulations, and many investigations have been devoted to this phenomena~\cite{MOA99,Sawford01,YB2004,BBC+05a,BOX+06,SYH08,SC09,TB09,BHB12,Eyink2011}.

In general, the separation $\langle R^2 (t) \rangle$ can be expressed as
\begin{equation}
\langle R^2 (t) \rangle = R_0^2 + 2 \int_0^t \int_0^{t^\prime} \langle \delta \mathbf{u} (t^\prime) \cdot \delta \mathbf{u} (t^{\prime\prime}) \rangle dt^{\prime\prime} dt^\prime ,
\label{eq:R2t}
\end{equation}
where $\delta \mathbf{u} (t) \equiv \mathbf{u}_1(t) - \mathbf{u}_2 (t)$ is the relative velocity between the two fluid particles. 
Equation \eqref{eq:R2t} thus relates the relative dispersion of two particles to
the Lagrangian correlation of relative velocity, 
$\langle \delta \mathbf{u}(0) \delta \mathbf{u} (\tau) \rangle$. In fact, 
considering $\langle \delta \mathbf{u}(0) \delta \mathbf{u} (\tau) \rangle$ in 
the limits of small values of $\tau$ ($\tau \rightarrow 0$), and
$\tau$ large compared to the velocity correlation time 
($\tau \rightarrow \infty$), leads to interesting information on the time-dependence of
$\langle R^2 (t) \rangle$. When $\tau \rightarrow \infty$, the relative 
velocity $\delta \mathbf{u}(\tau)$ becomes independent of its initial value 
$\delta \mathbf{u}(0)$, so the double integral in Eq.~\eqref{eq:R2t} is 
linear in $t$. This corresponds to a diffusion-like regime that is similar to 
the turbulent diffusion of single particles first discussed in the seminal
work of G.~I.~Taylor~\cite{Taylor:1922}. 
In the opposite limit of very short times, $\tau \rightarrow 0$, 
we can expand the integrand in Eq.~\eqref{eq:R2t} at $t=0$ in power series of $t$ and integrate to obtain:
\begin{equation}
\langle R^2(t) \rangle = R_0^2 + \langle \delta u^2 (0) \rangle t^2 + \langle \delta \mathbf{u} (0) \cdot \delta \mathbf{a} (0) \rangle t^3 + \mathcal{O}(t^4) ,
\label{eq:R2tseries}
\end{equation}
where $\delta u^2 (0)$ is a simplified notation for 
$\delta \mathbf{u} (0) \cdot \delta \mathbf{u} (0)$ 
and $\delta \mathbf{a} (t) \equiv \mathbf{a}_1 (t) - \mathbf{a}_2(t)$ 
is the relative acceleration between the two particles.
Equation \eqref{eq:R2tseries} shows that as long as the initial separation 
$R_0$ is non-zero, 
the initial velocity difference does not vanish and $\langle R^2 (t) \rangle$ 
is dominated by the $t^2$ term at very small times. 
The Richardson $t^3$ regime, therefore, can only exist for some intermediate time $t$ 
after marking the two particles~\cite{Batchelor:1950,Batchelor:1952a,MoninYaglom:v2}. 
Furthermore, we note that $\delta \mathbf{u}(t) \cdot \delta \mathbf{a}(t) = \frac{d}{dt} \Big[ \frac{1}{2} \delta u^2 (t) \Big]$ 
is the rate of change of the kinetic energy in the relative motion between the two particles. 
It has been shown that for separation $R_0$ in the inertial range, 
this rate of kinetic energy change is related to the turbulent energy cascade 
through scales~\cite{MOA99,FGV01,PSC01}:
\begin{equation}
\langle \delta \mathbf{u} \cdot \delta \mathbf{a} \rangle_{R_0} = \frac{d}{dt} \left\langle \frac{1}{2} \delta u^2 \right\rangle_{R_0} = - 2 \dissip,
\label{eq:duda}
\end{equation}
in which the averaging is taking over all particle pairs separated by a distance $R_0$ in the inertial range.
As we mentioned before, the energy flux $\dissip$ is positive for 3D flows. 
Hence in 3D turbulence the kinetic energy in the relative motion 
conditioned on a given separation $R_0$ between particles initially
\emph{decreases}. It increases at later times, consistent with 
the faster than $t^2$ increase of $\langle R^2(t) \rangle$ at later times.
This unexpected consequence of the energy cascade has been 
confirmed, both in numerical simulations~\cite{PSC01} and in 
Lagrangian particle tracking experiments~\cite{XOB08}. Note that substituting 
Eq.~\eqref{eq:duda} into Eq.~\eqref{eq:R2tseries} gives a negative $t^3$ term for $\langle R^2(t) \rangle$, 
which should not be confused with the positive coefficient in the Richardson dispersion law expected at later times.

In the problem of mixing of a passive scalar, a proper modeling of the 
fluctuations of concentration rests on understanding how two fluid particles
arrive at a given distance apart, or in other words,
how $\langle R^2(t) \rangle$ changes when $t<0$.
This amounts to tracking the motion of particles
backward in time. 
For backward dispersion, it is also expected that a Richardson-like regime exists, i.e., 
$\langle R^2(-t) \rangle \propto t^3$ for intermediate time $t$, 
but with a larger coefficient, which means that backward dispersion is faster than 
forward dispersion~\cite{SYB05,BLMO:2006,BIC:2014}. 
An interesting observation is that for incompressible Navier-Stokes turbulence, 
both Eq.~\eqref{eq:R2t} and Eq.~\eqref{eq:R2tseries} are also valid for $t<0$. 
We note that in 3D turbulence, the $t^3$ term in Eq.~\eqref{eq:R2tseries}
is {\it positive} for $t<0$, which implies that even for short times, 
backward dispersion is faster than forward dispersion.
This is a consequence of the fact that the kinetic energy of the
relative motion of particle pairs, followed backwards in time, {\it increases} 
at short times, contrary to what happens when following particle motion forward 
in time.

This property is a manifestation of time irreversibility of turbulence.
In principle, it can be used to detect the ``arrow-of-time'', while following many particle trajectories in a turbulent flow~\cite{JXPB14}.
This manifestation of time-irreversibility in the relative dispersion 
between two fluid particles, $d \langle R^2 \rangle / dt$, ultimately rests
 on the relation between the turbulent energy cascade and the rate of energy 
change in the relative motion expressed by Eq.~\eqref{eq:duda}. 
This means that if we know the derivative of the kinetic energy in the 
relative motion, $d(\frac{1}{2} \delta u^2 )/dt$, or the velocity field and 
its time derivative $d\delta \mathbf{u}/dt$, then we can also detect the ``arrow of time''. 
On the same basis, other Lagrangian quantities that combine 
both relative velocity and separation $R$ can also be formed with the property that 
their time derivatives are sensitive to whether the ``arrow of time'' is flipped or not~\cite{FF13}. 
A further observation is that for other flows that do not satisfy incompressible Navier-Stokes equations, 
Eq.~\eqref{eq:duda} might not remain the same for $t>0$ and $t<0$, i.e., 
there could be an anomaly in Lagrangian velocity statistics. 
For example, it has been shown that for the compressible Burgers equation, 
taking $d (\delta u^2) / dt$ at $t=0$ from the $t<0$ and $t>0$ side give different values, 
which is due to the formation of shocks in Burgers turbulence when time is running forward~\cite{FF14}. 
This is clearly a stronger manifestation of time irreversibility.

While Eq.~\eqref{eq:duda} has been derived for both 2D and 3D turbulence, 
its validity has so far been verified numerically and experimentally only
for 3D turbulence, but not for 2D turbulence. 
The physics of energy cascade is completely different 
in 2D turbulence~\cite{K67,FGV01,BE12}, compared to that in 3D turbulence. 
In particular, the energy flux in 2D is towards larger scales,
and consequently the kinetic energy in the relative motion between fluid particle pairs 
is expected to increase initially. Confronting this prediction with numerical and experimental data 
could be an interesting and important work for the future.

\subsection{Shape deformation and structure of the flow}

While the previous section was devoted to the relative motion between
two particles, we now turn to the Lagrangian statistics involving more than two particles. 
The description of a set of points requires not only a size, such as 
the mean distance between the particles, but also extra variables describing
the shape of the set of points.
The shape evolution provides interesting information on the local (topological) 
structure of the turbulent flow, which cannot be obtained from the study of the
mean separation between pairs of particles alone. To explore the 
flow topology, one needs to follow at least 3 particles in a 2D flow and 4 particles in a 3D flow. 
It has been observed that the evolution of initially isotropic objects (equilateral triangles or regular tetrahedra) 
in turbulent flows differs from that in a Gaussian velocity field.
Qualitatively, the shapes obtained in a turbulent flow
are more elongated at intermediate times than expected by using a
flow with Gaussian statistics, or 
before all particles are widely separated so their velocities become 
independent~\cite{PSC00,CP2001,BBC+05b,XOB08,HYS:2011}. 
To see how this is related to flow topology, one can define an effective local velocity gradient 
$\mathbf{M}$ perceived by the set of particles~\cite{CPS99,PSC01,XPB11,PBX13}:
\begin{equation}
\mathbf{M} = \mathbf{g}^{-1} \mathbf{W},
\label{eq:perceivedVG}
\end{equation}
where the matrices $\mathbf{g}$ and $\mathbf{W}$ are defined as
\begin{equation}
g_{ij} = \sum_{\alpha = 1}^{N} x^{\prime \alpha}_i x_j^{\prime \alpha}
\end{equation}
and
\begin{equation}
W_{ij} = \sum_{\alpha = 1}^{N} x^{\prime \alpha}_i u_j^{\prime \alpha}
\end{equation}
where $N$ is the total number of particles in the set used to define the perceived velocity gradient $\mathbf{M}$, 
\begin{equation}
\mathbf{x}^{\prime \alpha} = \mathbf{x}^{\alpha} - \frac{1}{N} \sum_{\alpha=1}^{N} \mathbf{x}^{\alpha}
\end{equation}
and
\begin{equation}
\mathbf{u}^{\prime \alpha} = \mathbf{u}^{\alpha} - \frac{1}{N} \sum_{\alpha=1}^{N} \mathbf{u}^{\alpha}
\end{equation}
are the position and velocity of particle $\alpha$ relative to the center of the particle set. It is easy to show that the perceived velocity gradient $\mathbf{M}$ given by Eq.~\eqref{eq:perceivedVG} is the least square fit of a linear velocity field from the velocities at the particle positions. When the separations between the particles are very small, in the range where viscous effects dominate, the perceived velocity gradient $\mathbf{M}$ given by Eq.~\eqref{eq:perceivedVG} recovers the true velocity gradient. When the particle separations are in the inertial range, $\mathbf{M}$ probes the inertial range dynamics, which is the main motivation to study $\mathbf{M}$~\cite{CPS99}. Other effective local velocity gradients similar to $\mathbf{M}$ have been proposed and the information on flow topology obtained are also comparable~\cite{LTK05,LOBM07}.

To probe the flow topology, it is helpful to decompose the velocity gradient 
$\mathbf{M}$ as a sum of a 
symmetrical part, $\mathbf{S} = (\mathbf{M} + \mathbf{M}^T)/2$, 
which represents 
the rate of strain (local stretching or compression) of the flow, 
and an anti-symmetrical part,
$\mathbf{\Omega} = (\mathbf{M} - \mathbf{M}^T)/2$, 
which represents the local rotation (by construction, $\mathbf{S} + \mathbf{\Omega} = \mathbf{M}$). 
This decomposition is in fact unique.
The strain $\mathbf{S}$ and the rotation $\mathbf{\Omega}$ interact with each other, 
which forms the rich dynamics of turbulent flows. In particular, 
on average the local rotation rate is constantly amplified because 
of the action of the strain, 
a phenomenon called ``vortex stretching'', which is eventually compensated by 
the viscous dissipation. 
Early studies of the true velocity gradient in turbulent flows have revealed that 
the stretching of the vorticity is closely tied to the statistics of the eigenvalues of the strain $\mathbf{S}$. 
Namely, among the three eigenvalues, which are all real 
because $\mathbf{S}$ is symmetric by definition,
the intermediate eigenvalue is predominately positive~\cite{Betchov:1956,Siggia:1981a}. 
This has been verified in numerical simulations and experiments~\cite{Kerr:1987,TKD1992} 
and has stimulated further studies on the dynamics of velocity gradients~\cite{Vieillefosse:1984,Cantwell:1992,Meneveau:2011}. 
A natural expectation from vortex stretching is that the vorticity vector would be 
preferentially aligned with the direction corresponding to the largest eigenvalue of the strain, 
which represents the strongest stretching. On the other hand, numerical and experimental data show 
that at any given instant, vorticity is preferentially aligned with the intermediate eigenvalue of the strain, 
which corresponds to rather mild stretching~\cite{Siggia:1981b,AKKG87,TKD1992}. 
This interesting observation has been studied extensively in subsequent research
(see \cite{Tsinober:book} for a detailed discussion).

When the size of the particle cluster used to obtain $\mathbf{M}$ from Eq.~\eqref{eq:perceivedVG} 
is larger than the viscous range, $\mathbf{M}$ differs from the true velocity gradient and 
provides a way to probe the flow property in the inertial range of scales. 
It has been observed that $\mathbf{M}$ obtained in this way shares 
qualitatively many properties of the true velocity gradient. In particular, 
the intermediate eigenvalue of the rate of strain $\mathbf{S}$ is predominately positive and instantaneously 
the vorticity is aligned with this intermediate eigenvalue~\cite{CPS99,PN12,PBX13}. 
These properties of $\mathbf{M}$, which are closely related to the 
inertial range dynamics and hence to energy cascade, are expected
to lead to irreversibility in Lagrangian multi-particle statistics. 
For example, the short-time deformation of an initially isotropic tetrahedron 
formed by four fluid particles is governed by the eigenvalues of the perceived rate of strain. 
Therefore, the non-zero average of the intermediate eigenvalue of $\mathbf{S}$ implies that 
the shape evolution of a tetrahedron differs when followed forward or backward in time~\cite{JXPB14}. 
The Lagrangian view also shows that the perceived vorticity vector indeed tends to 
align with the largest eigenvalue of $\mathbf{S}$, in the sense that the vorticity vector 
turns to the initially strongest stretching direction, but with a time delay such that 
at any given instant the vorticity is observed to preferentially align with the intermediate eigenvalue~\cite{XPB11}. 
This property is also found for the true velocity gradient~\cite{CM2011,PBX13}.
The observed alignment process of vorticity with the rate of strain at a given time will completely differ when following 
flow trajectories backward in time.
We note that when the length $R_0$ characteristic of the set of points is in
the inertial range of scales, the dynamical processes are essentially
self-similar. This can be seen by rescaling time by the time scale 
$t_0 = (R_0^2/\dissip)^{1/3}$, which can be viewed as the eddy-turnover time
at scale $R_0$~\cite{XPB11,PBX13,JXPB14}.

A different but related interesting question is how rigid particles with given 
shapes see the turbulent flow. For small rod-like particles, it is surprising to 
observe that they align with local vorticity and hence preferentially with the 
intermediate eigenvector of the rate of strain~\cite{PW11,NOV14,NKOV2015}. 
New results are available concerning the coupling between translation 
and rotation of neutrally buoyant particles with other shapes, such
as large spheres~\cite{ZGB+11,KGBB13}, 
ellipsoids~\cite{BV12,CM13,Gust+14}, or other anisotropic shapes~\cite{MPK+14}.
It would be very interesting to find how their dynamics are related to the irreversibility of the flow.

In summary, we note that for multi-particle Lagrangian statistics, 
the distance between particles defines a natural length scale of the problem and 
the energy cascade process in turbulence inherently causes the observed irreversibility. 
Stochastic models have been widely used to describe the multi-particle dispersion process and 
many aspects of the observed multi-particle statistics can be recovered by these models~\cite{Devenish:2013,SPY2013}. 
As almost all these models have time-reversible dynamics built in, 
one should be cautious not to push these models beyond the range in which they are valid.

\section{Single-particle statistics}

In situations involving several particles, we can naturally
introduce one length scale (or more) in the problem, 
therefore permitting to establish a relation with the Eulerian correlation 
functions of the velocity field, and hence the energy flux or energy 
dissipation. In contrast, following only one 
fluid particle in a turbulent flow does not give rise to an unambiguous
identification of a length scale.
For this reason, new ideas and concepts are needed in order to understand
the flow properties from the statistics of single particle trajectories only.
Recent progress provides new insights into these interesting questions.

\subsection{Velocity structure functions}
\label{sec:single}

The most studied single-particle statistic quantities are the Lagrangian velocity structure functions, i.e., the moments of the velocity increments following a fluid particle:
\begin{equation}
S_n (\tau) = \langle (\delta_\tau u)^n \rangle \equiv \langle [u(t+\tau) - u(t)]^n \rangle,
\label{eq:Sn}
\end{equation}
where $u(t)$ is one component of the particle velocity at time $t$ 
along a direction, being understood that 
for homogeneous and isotropic turbulence, the choice of the component does not matter. 
By analogy with the Eulerian velocity increments, one may surmise that
the Lagrangian velocity increments $\delta_\tau u$ depend on the turbulent 
energy dissipation rate $\dissip$. Furthermore, if the time lag $\tau$ is much 
larger than the viscous time scale but smaller than the largest time scale of 
the flow, it is tempting to postulate that the statistics of the velocity 
increments are universal and independent of viscosity. Simple dimensional 
analysis then leads to the scaling $\delta_\tau u \sim (\dissip \tau)^{1/2}$ 
and hence $S_n (\tau) \sim (\dissip \tau)^{n/2}$~\cite{Landau_FM,MoninYaglom:v2,Yeung:2002,TB09}. 
Available experimental and numerical data show that the dependence
of $S_n(\tau)$ on $\tau$ has very little to do with
the expected scaling behavior~\cite{MMMP01,BBC+04,XBOB06}. Various theories have been proposed to explain the observed deviations~\cite{TB09,ZS:2010,He:2011}, with the multifractal model being the most popular~\cite{Borgas:1993,CRL+03,BBC+04} (see also a recent summary in Ref.~\cite{CCA+12}).

Among the Lagrangian structure functions, the second order, obtained by
taking $n=2$ in 
Eq.~\eqref{eq:Sn}, is of special interest because according to the dimensional 
argument it is proportional to the energy dissipation rate $\dissip$ 
itself, so the average in Eq.~\eqref{eq:Sn} is not affected by the
strong fluctuations in $\dissip$, which is 
known to lead to corrections to scaling in the case of the spatial structure 
functions (intermittency corrections)~\cite{K62,MoninYaglom:v2,frisch:1995}.
Based on these considerations, the scaling 
$S_2(\tau) \propto \dissip \tau$ is expected to be exact, just as the 
linear scaling of the third-order Eulerian velocity structure function 
predicted by Eq.~\eqref{eq:45law} (the 4/5-law). 
This expectation is summarized by the following relation:
\begin{equation}
S_2 (\tau) = \langle (\delta_\tau u)^2 \rangle = C_0 \dissip \tau,
\label{eq:C0}
\end{equation} 
where $C_0$ is expected to be a universal constant of order 
unity~\cite{MoninYaglom:v2,Yeung:2002,TB09}. 
The best available data, from
state-of-the-art experiments and numerical simulations, however, does 
{\it not} support the scaling suggested by Eq.~\eqref{eq:C0}~\cite{TB09,SY11}. 
If anything, the values of $C_0$ observed for 3D turbulence are found to 
increase with the Reynolds number of the flow and are approximately $7$ for the highest 
Reynolds numbers measured so far~\cite{TB09,SY11}. 
In the case of 2D turbulence, where Eq.~(\ref{eq:C0})
is also expected to hold,
the values of $C_0$ increase much 
faster with the Reynolds number. The results of numerical simulations
at the highest available resolution suggest values of the order
$\sim \mathbf{O}(10^2)$, without any indication of saturation~\cite{FXP+13}. 
This casts a serious doubt on the scaling predicted by using dimensional
arguments.

In fact, the assumption that the statistics of velocity increments following a 
fluid particle, $\delta_\tau u$, depend on $\dissip$, the energy flux through 
spatial scales, is questionable. That assumption is directly inspired from
Kolmogorov's hypotheses on the statistics of Eulerian velocity increments. 
As already noticed, establishing
a connection between Lagrangian and Eulerian statistics requires the 
introduction of a length scale into the structure functions $S_n(\tau)$, 
which is achieved by assuming that
$\tau \delta_\tau u $ is equivalent to the separation $r$ in Eulerian 
statistics and that $\delta_\tau u$ scales as the Eulerian velocity 
difference $\delta_r u$. Substituting $\delta_r u \sim \delta_\tau u$ and 
$r \sim \tau \delta_\tau u$ into Eq.~\eqref{eq:45law} 
leads to $\langle (\delta_\tau u)^2 \rangle \sim \dissip \tau$, hence to
a formal justification of Eq.~\eqref{eq:C0}.

The formal analogy between the Eulerian velocity increments, $\delta_r u$,
and the Lagrangian velocity increments $\delta_\tau u$, through the use
of $r \approx \tau \delta_\tau u$ should, however, be taken very carefully.
The Eulerian statistics of $\delta_r u$ are 
mostly determined by turbulent eddies of size $r$. 
When estimating the Lagrangian increment $\delta_\tau u$ using the
Eulerian increment $\delta_r u$ with $r \approx \tau \delta_\tau u$, it should be 
noticed that the time $\tau$ necessary for a particle to travel up to $r$,
$\tau \sim r/\delta_r u \sim (\delta_r u)^2/\dissip$, is in fact
the life time of an eddy of size $r$. That is to say, for a 
fluid particle to move a distance $r$, its velocities at the start, $u(t)$, and 
at the end, $u(t+\tau)$, are unlikely to be the result of the same eddy 
of size $r$.
This essential dissimilarity between $\delta_r u$ and $\delta_\tau u$ leads to
very different properties between Eulerian and Lagrangian statistics. More 
generally, this feature highlights
statistically stationary turbulence, as an 
ultimate example of non-equilibrium steady state, far from 
equilibrium~\cite{RS78}.

These, and other considerations led Falkovich et al.~\cite{FXP+13} to 
question the validity of Eq.~\eqref{eq:C0}. 
They pointed out that the statistics of $\delta_\tau u$, including all 
Lagrangian velocity structure functions, are symmetric under the transformation 
of $t \rightarrow -t$, therefore being unable to pick up the fundamental
time-irreversibility of the flow.
Therefore, there is no fundamental reason to relate the statistics of
$\delta_\tau u$ to the energy flux, which is the cause of the 
time-irreversibility of turbulent flows.

In summary, there is strong motivation to consider other Lagrangian statistics that reveal the irreversible nature of turbulent flows. That is the topic we cover in the next subsection.

\subsection{Kinetic energy increments and instantaneous power}

An interesting recent discovery is that the change of kinetic energy following individual fluid particles can be used to detect the ``arrow of time''. In Ref.~\cite{XPFB14}, it was observed in experiments and numerical simulations that the third moments of the kinetic energy change, 
$\delta_\tau W \equiv [u^2(\tau) - u^2(0)]/2$:
\begin{equation}
\langle (\delta_\tau W)^3  \rangle  \equiv \langle [u^2(t+\tau)/2 - u^2(t)/2]^3\rangle
\end{equation}
are negative for time lags $\tau$ positive, but 
smaller than the velocity correlation time (the largest time scale of the 
flow). 
This implies that the probability distribution of the instantaneous power, 
$p = \lim_{\tau \rightarrow 0} \delta_\tau W / \tau = \mathbf{u} \cdot \mathbf{a}$, 
is negatively skewed. The origin of this skewness can be traced back 
to the observed tendency of
fluid particles to gain kinetic energy slowly, but lose it more
suddenly. This provides a way to identify the arrow of time, as flipping
$t \rightarrow -t$ would lead to the exact opposite: particles would
gain energy faster than they dissipate it.
The negative skewness of the distribution of $p$ was observed for both 2D and 
3D turbulence (at least for 2D turbulent flows that were agitated with 
forces short-correlated in time), i.e., independent of whether energy flows towards
larger or smaller scales. 
From the more general point of view of energy exchange, 
kinetic energy is dissipated into heat in an irreversible way in turbulent
flows, both in 2D and 3D.
In that sense, the qualitative similarity between the statistics of $p$ 
and $\delta_\tau W$ in both 2D and 3D turbulences, once expressed
in terms of the energy dissipation, $\dissip$, may not be so surprising.
Available data support that the moments of $p / \dissip$ follow, to a
good approximation, a power law dependence on 
the Reynolds number of the flow, 
with an exponent independent of the spatial dimension.
This suggests that the third moment 
$- \langle p^3 \rangle / \dissip^3$ can be used as a measure of irreversibility.
Moreover, the scaling of the third moment $-\langle p^3 \rangle / \dissip^3$ 
can be qualitatively explained by assuming that
$-\langle p^3 \rangle / \dissip^3$ is dominated by the extreme events of 
negative $p$ with large magnitudes, i.e., events when fluid particles lose 
kinetic energy very rapidly, an argument pictorially alluding to 
``flight-and-crash'' events~\cite{XPFB14}.
This skewed distribution of $p$ also manifests itself in the negative skewness of the kinetic energy change associated with single velocity component in a 3D turbulent flow~\cite{Mordant:thesis}, and in the negative skewness of the longitudinal Lagrangian velocity increments~\cite{LN2014}. 

While the skewness of the instantaneous power $p$ is negative for both 2D and 
3D turbulence, and in this sense, seems to be insensitive to the very different
physical mechanisms of cascade in these two cases, one may nevertheless ask 
which quantity reflects the difference in the dynamics in 2D and 3D flows. 
To answer that question, one can decompose the instantaneous power $p$ into
\begin{equation}
p = \mathbf{u}\cdot\mathbf{a} = - \mathbf{u} \cdot \nabla P + \mathbf{u} \cdot \mathbf{f} + \mathbf{u} \cdot \mathbf{D},
\label{eq:powerDecomp}
\end{equation}
where $-\nabla P$, $\mathbf{f}$, and $\mathbf{D}$ are the pressure gradient, external forces, and dissipative forces, respectively. In 3D flows, the dissipative forces consist of the viscous forces alone, $\mathbf{D} = \nu \nabla^2 \mathbf{u}$; while for 2D flows, the dissipative forces include both viscous forces and friction forces, $\mathbf{D} = \nu \nabla^2 \mathbf{u} - \alpha \mathbf{u}$, where $\alpha > 0$ is the linear friction coefficient. Numerical simulation data show~\cite{PXB+14} that in both 2D and 3D flows, the magnitude of the pressure gradient term $- \mathbf{u} \cdot \nabla P$ overwhelms all other terms and determines the magnitude of $p$, but the contributions to the third moment of $p$ are more subtle and show interesting differences between 2D and 3D flows. 

In 2D flows, the pressure gradient term is also negatively skewed, 
and it contributes to nearly $2/3$ of $\langle p^3 \rangle$, 
with the other dominant contribution being provided by the correlation between
the pressure gradient and the friction, $\langle (-\mathbf{u} \cdot \nabla P)^2 (- \alpha \mathbf{u} \cdot \mathbf{u}) \rangle$. 
In 3D flows, the 
situation is completely different: the skewness of the pressure gradient term is very small, 
even slightly positive, so its direct contribution to 
$\langle p^3 \rangle$ is very small and of opposite sign, compared to 
$\langle p^3 \rangle$. The dominant term that contributes to 
$\langle p^3 \rangle$ is the cross term between the pressure gradient and the 
viscous forces, $\langle (- \mathbf{u} \cdot \nabla P)^2 (\nu \mathbf{u} \cdot \nabla^2 \mathbf{u}) \rangle$~\cite{PXB+14}.

Therefore, the pressure gradient term acts very differently in 2D and 3D flows.
In 2D, it behaves according to naive expectation, insofar as it provides
the dominant term of the fluctuations, and contributes significantly to the
observed asymmetry of the distribution of $p$. In comparison,
although the pressure term in 3D also provides the main contribution to the variance
of $p$, it hardly provides any significant contribution to the observed
asymmetry of the distribution of $p$.

Note that in homogenous flows, the pressure gradient term alone does not 
change the total energy in the flow since 
$\langle \mathbf{u} \cdot \nabla P \rangle = 0$, i.e., the pressure gradient 
term merely redistributes kinetic energy within the flow. The different role 
played by the pressure gradient term would imply that the way of energy redistribution 
is different in 2D and 3D flows. 
Indeed, the averaged value of $-\mathbf{u} \cdot \nabla P$ conditioned
on the kinetic energy of the particles reveals that
in 2D flows, particles get as much energy from pressure gradient forces as they
lose it, independently of their velocity:
$\langle - \mathbf{u} \cdot \nabla P | \mathbf{u}^2 \rangle = 0$ for all $\mathbf{u}^2$. 
In contrast, in 3D flows, the mean pressure contribution 
conditioned on the velocity is negative for particles with small 
velocities: $\langle - \mathbf{u} \cdot \nabla P | \mathbf{u}^2 \rangle < 0$ for $\mathbf{u}^2 \lesssim 2 \langle \mathbf{u}^2 \rangle$, and it is positive for particles with large velocities. This implies that the pressure gradient term takes kinetic energy away from slow particles and gives it to fast particles~\cite{PXB+14}. 
Without other terms to stop this action, the pressure gradient term 
alone could potentially drive the flow into singularities. This 
observation might provide new insight into the long-standing Millennium problem
on the regularity of the Navier-Stokes equations~\cite{Leray:1934,Fefferman:2006}.

The decomposition of the instantaneous power $p$ in the form of 
Eq.~\eqref{eq:powerDecomp} is certainly not unique. A possible 
alternative consists in decomposing the fluid acceleration into a local 
part $\mathbf{a}_L = \partial \mathbf{u} / \partial t$ and a convective part 
$\mathbf{a}_C = \mathbf{u} \cdot \nabla \mathbf{u}$. This decomposition 
separates the effect of the flow seen by
the particle as resulting from the time variation of the velocity
field locally, $\mathbf{a}_L$, 
and from the advection by a time-independent (frozen) flow, 
$\mathbf{a}_C$.
It has been noticed that the two components $\mathbf{a}_L$ and 
$\mathbf{a}_C$ cancel each other to a large 
extent~\cite{TVY2001,gulitski:2007b}. How their contributions to the instantaneous power $p$ behave and what they reveal about the irreversibility of the flow are interesting problems for future study.

The results discussed above are all for incompressible turbulence. As predicted by the study of the two-particle statistics governed by Burgers equation~\cite{FF14}, in compressible flows, the irreversibility is expected to manifest itself in a stronger way. It would be interesting to confirm these predictions, by either experimental or numerical simulation data.

Another direction worth exploring concerns the effect of particle inertia 
on the effects discussed here for fluid particles. Particles whose densities 
differ from the fluid density or whose sizes are larger than the Kolmogorov scale do 
not follow the flow faithfully due to their inertia. There has been a 
wealth of literature on the dynamics of various inertia 
particles~\cite{voth:2002,BBB+06,QBB+07,VMVP08,XB08,ZFG+13}. It will be 
therefore interesting to carry out analyses similar to that in Refs.~\cite{XPFB14,PXB+14} to see how the irreversibility of the flow is reflected in the dynamics of inertial particles.

\section{Summary and discussion}

In this brief review, we discussed 
how Lagrangian statistics, obtained by following the motion of fluid particles in the flow,
are sensitive to the intrinsic time irreversibility of turbulence.
This irreversibility is a consequence of the property that, in
turbulent flows, the 
kinetic energy is supplied into the fluid motion at a scale very different from the scale 
at which the energy is dissipated. In high-Reynolds number flows, where the 
forcing scales and the dissipation scales are widely separated, 
energy is transferred at a constant rate through scales, over a wide 
range of scales.
This energy cascade and dissipation process 
are irreversible and the issue is how these phenomena affect the motion of fluid particles.

We emphasize that depending on the spatial dimension, the energy is 
transferred either to small scales or to large scales, to be dissipated
either by viscosity (in 3D), or by friction (in 2D). This results in very
different physical mechanisms.
The investigation of multi-particle Lagrangian statistics, obtained by 
following sets of particles, allows the identification of at least one length scale. 
As a consequence, multi-particle Lagrangian statistics
naturally sense the energy flux and 
reflect it in the change of kinetic energy associated with the relative motion,
the dynamics of the perceived velocity gradients and the shape deformation of 
isotropic objects. Single-particle statistics, on the other hand, 
does not permit such an unambiguous identification of a length scale. 
Previous attempts to construct a length scale from single particle 
trajectories and hence to connect single-particle statistics with Eulerian 
statistics that depend on the energy flux are fundamentally 
questionable, and lead to incorrect predictions. While the statistics of
the usual Lagrangian structure function do not permit to distinguish
the arrow of time, the
statistics of other quantities, such as the energy change following a 
fluid particle, do reveal the irreversibility of the flow. In this case,
irreversibility is due to the energy dissipation, i.e., kinetic energy is eventually converted to thermal energy. Those statistics, therefore, show similar behavior in both 2D and 3D, despite the opposite directions of the energy flux in flows with different spatial dimensions.

Most of the work reviewed here originated from observations in physical 
experiments and numerical simulations that built on rapid progresses
in experimental 
techniques and computational power. As more and more high-quality, 
high-resolution data are generated by the research 
community and are being shared with the whole community 
(e.g., in large public databases~\cite{li:2008,yu:2012}), we expect that other properties of 
Lagrangian statistics in turbulent flows will be discovered and their
connection with the time irreversibility of the flow, or more generally, 
with the dynamical properties of the flow, be understood. From a broader 
theoretical 
point of view, fluid turbulence is a well-known example of out-of-equilibrium 
system. How to relate what we learned from studying fluid turbulence to other 
non-equilibrium systems is another open area for future investigation.

\begin{acknowledgments}
We thank G. Boffetta, G. Falkovich, N. Francois, R. Grauer, J. Jucha, M. Shats, and H. Xia
for stimulating discussions during our work.
We are grateful to the Max Planck Society for continuous support to our research.
AP also acknowledges financial support from ANR (contract TEC 2), 
the Alexander von Humboldt Foundation, 
and the PSMN at the Ecole Normale Sup\'erieure de Lyon.
\end{acknowledgments}

%\bibliographystyle{elsart-num}
%\bibliography{./turb_refs}

\begin{thebibliography}{99}

\bibitem{Landau_FM}
L.~D. Landau and E.~M. Lifshitz.
\newblock {\em Fluid Mechanics}.
\newblock Butterworth-Heinemann, 1987.

\bibitem{Landau_SP}
L.~D. Landau and E.~M. Lifshitz.
\newblock {\em Statistical Physics}.
\newblock Butterworth-Heinemann, 1980.

\bibitem{Derrida:2007}
B.~Derrida.
\newblock Non-equilibrium steady states: fluctuations and large deviations of
  the density and of the current.
\newblock {\em J. Stat. Mech. Theor. Exp.}, 9:P07023, 2007.

\bibitem{CJP10}
S.~Ciliberto, S.~Joubaud, and A.~Petrosyan.
\newblock Fluctuations in out-of-equilibrium systems: from theory to
  experiment.
\newblock {\em J. Stat. Mech.}, 12:P12003, 2010.

\bibitem{TennekesLumley:1972}
T.~Tennekes and J.~L. Lumley.
\newblock {\em A First Course in Turbulence}.
\newblock The MIT Press, Cambridge, USA, 1972.

\bibitem{frisch:1995}
U.~Frisch.
\newblock {\em Turbulence: The Legacy of A.~N.~Kolmogorov}.
\newblock Cambridge University Press, Cambridge, England, 1995.

\bibitem{pope:2000}
S.~B. Pope.
\newblock {\em Turbulent Flows}.
\newblock Cambridge University Press, Cambridge, England, 2000.

\bibitem{K67}
R.~H. Kraichnan.
\newblock Inertial ranges in two-dimensional turbulence.
\newblock {\em Phys. Fluids}, 10:1417--1423, 1967.

\bibitem{BE12}
G.~Boffetta and R.~E. Ecke.
\newblock Two-dimensional turbulence.
\newblock {\em Annu. Rev. Fluid Mech.}, 44:417--451, 2012.

\bibitem{RS78}
H.~A. Rose and P.~L. Sulem.
\newblock Fully developed turbulence and statistical mechanics.
\newblock {\em J. Physique}, 39:441--484, 1978.

\bibitem{K41a}
A.~N. Kolmogorov.
\newblock The local structure of turbulence in incompressible viscous fluid for
  very large {R}eynolds numbers.
\newblock {\em Dokl. Akad. Nauk SSSR}, 30:301--305, 1941.

\bibitem{K41c}
A.~N. Kolmogorov.
\newblock Dissipation of energy in the locally isotropic turbulence.
\newblock {\em Dokl. Akad. Nauk SSSR}, 32:16--18, 1941.

\bibitem{FGV01}
G.~Falkovich, K.~Gawedzki, and M.~Vergassola.
\newblock Particles and fields in fluid turbulence.
\newblock {\em Rev. Mod. Phys.}, 73:913--975, 2001.

\bibitem{MOA99}
J.~Mann, S.~Ott, and J.~S. Andersen.
\newblock Experimental study of relative, turbulent diffusion.
\newblock Technical Report Ris\o-R-1036(EN), Ris\o~National Laboratory, 1999.

\bibitem{laPorta01}
A.~La~Porta, G.~A. Voth, A.~M. Crawford, J.~Alexander, and E.~Bodenschatz.
\newblock Fluid particle accelerations in fully developed turbulence.
\newblock {\em Nature}, 409:1017--1019, 2001.

\bibitem{MMMP01}
N.~Mordant, P.~Metz, O.~Michel, and J.-F. Pinton.
\newblock Measurement of {Lagrangian} velocity in fully developed turbulence.
\newblock {\em Phys. Rev. Lett.}, 87:214501, 2001.

\bibitem{LTK05}
B.~L\"uthi, A.~Tsinober, and W.~Kinzelbach.
\newblock Lagrangian measurements of vorticity dynamics in turbulent flow.
\newblock {\em J. Fluid Mech.}, 528:87--118, 2005.

\bibitem{BOX+06}
M.~Bourgoin, N.~T. Ouellette, H.~Xu, J.~Berg, and E.~Bodenschatz.
\newblock The role of pair dispersion in turbulent flow.
\newblock {\em Science}, 311:835--838, 2006.

\bibitem{VMVP08}
R.~Volk, N.~Mordant, G.~Verhille, and J.-F. Pinton.
\newblock Laser doppler measurement of inertial particle and bubble
  accelerations in turbulence.
\newblock {\em Europhys.~Lett.}, 81:34002, 2008.

\bibitem{XPB11}
H.~Xu, A.~Pumir, and E.~Bodenschatz.
\newblock The pirouette effect in turbulent flows.
\newblock {\em Nature Phys.}, 7:709--712, 2011.

\bibitem{YP89}
P.~K. Yeung and S.~B. Pope.
\newblock Lagrangian statistics from direct numerical simulations of isotropic
  turbulence.
\newblock {\em J. Fluid Mech.}, 207:531--586, 1989.

\bibitem{PSC00}
A.~Pumir, B.~I. Shraiman, and M.~Chertkov.
\newblock Geometry of {Lagrangian} dispersion in turbulence.
\newblock {\em Phys. Rev. Lett.}, 85:5324--5327, 2000.

\bibitem{YB2004}
P.~K. Yeung and M.~S. Borgas.
\newblock Relative dispersion in isotropic turbulence. {Part} 1. {Direct}
  numerical simulations and {Reynolds}-number dependence.
\newblock {\em J. Fluid Mech.}, 503:93--124, 2004.

\bibitem{BBC+05b}
L.~Biferale, G.~Boffetta, A.~Celani, B.~J. Devenish, A.~Lanotte, and F.~Toschi.
\newblock Multiparticle dispersion in fully developed turbulence.
\newblock {\em Phys. Fluids}, 17:111701, 2005.

\bibitem{HYS:2011}
J.~F. Hackl, P.~K. Yeung, and B.~L. Sawford.
\newblock Multi-particle and tetrad statistics in numerical simulations of
  turbulent dispersion.
\newblock {\em Phys. Fluids}, 23:065103, 2011.

\bibitem{FXP+13}
G.~Falkovich, H.~Xu, A.~Pumir, E.~Bodenschatz, L.~Biferale, G.~Boffetta, A.~S.
  Lanotte, and F.~Toschi.
\newblock On lagrangian single-particle statistics.
\newblock {\em Phys. Fluids}, 24:055102, 2012.

\bibitem{Richardson1926}
L.~F. Richardson.
\newblock Atmospheric diffusion shown on a distance-neighbour graph.
\newblock {\em Proc. Roy. Soc. Lond. A}, 110:709--737, 1926.

\bibitem{Sawford01}
B.~L. Sawford.
\newblock Turbulent relative dispersion.
\newblock {\em Annu. Rev. Fluid Mech.}, 33:289--317, 2001.

\bibitem{BBC+05a}
L.~Biferale, G.~Boffetta, A.~Celani, B.~J. Devenish, A.~Lanotte, and F.~Toschi.
\newblock Lagrangian statistics of particle pairs in homogeneous isotropic
  turbulence.
\newblock {\em Phys. Fluids}, 17:115101, 2005.

\bibitem{SYH08}
B.~L. Sawford, P.~K. Yeung, and J.~F. Hackl.
\newblock {Reynolds} number dependence of relative dispersion statistics in
  isotropic turbulence.
\newblock {\em Phys. Fluids}, 19:065111, 2008.

\bibitem{SC09}
J.~P. L.~C. Salazar and L.~R. Collins.
\newblock Two-particle dispersion in isotropic turbulent flows.
\newblock {\em Annu.~Rev.~Fluid Mech.}, 41:405--432, 2009.

\bibitem{TB09}
F.~Toschi and E.~Bodenschatz.
\newblock Lagrangian properties of particles in turbulence.
\newblock {\em Annu. Rev. Fluid Mech.}, 41:375--404, 2009.

\bibitem{BHB12}
R.~Bitane, H.~Homman, and J.~Bec.
\newblock Time scales of turbulent relative dispersion.
\newblock {\em Phys. Rev. E}, 86:045302, 2012.

\bibitem{Eyink2011}
G.~L. Eyink.
\newblock Stochastic flux freezing and magnetic dynamo.
\newblock {\em Phys. Rev. E}, 83:056405, 2011.

\bibitem{Taylor:1922}
G.~I. Taylor.
\newblock Diffusion by continuous movements.
\newblock {\em Proc. Lond. Math. Soc.}, 20:196--212, 1922.

\bibitem{Batchelor:1950}
G.~K. Batchelor.
\newblock The application of the similarity theory of turbulence to atmospheric
  diffusion.
\newblock {\em Q. J. R. Meteorol. Soc.}, 76:133--146, 1950.

\bibitem{Batchelor:1952a}
G.~K. Batchelor.
\newblock Diffusion in a field of homogeneous turbulence.~{II}.~{The} relative
  motion of particles.
\newblock {\em Proc. Camb. Phil. Soc.}, 48:345--362, 1952.

\bibitem{MoninYaglom:v2}
A.~S. Monin and A.~M. Yaglom.
\newblock {\em Statistical Fluid Mechanics}, volume~2.
\newblock MIT Press, Cambridge, MA, 1975.

\bibitem{PSC01}
A.~Pumir, B.~I. Shraiman, and M.~Chertkov.
\newblock The {Lagrangian} view of energy transfer in turbulent flow.
\newblock {\em Europhys. Lett.}, 56:379--385, 2001.

\bibitem{XOB08}
H.~Xu, N.~T. Ouellette, and E.~Bodenschatz.
\newblock Evolution of geometric structures in intense turbulence.
\newblock {\em New J. Phys.}, 10:013012, 2008.

\bibitem{SYB05}
B.~L. Sawford, P.~K. Yeung, and M.~S. Borgas.
\newblock Comparison of backwards and forwards relative dispersion in
  turbulence.
\newblock {\em Phys. Fluids}, 17:095109, 2005.

\bibitem{BLMO:2006}
J.~Berg, B.~L\"uthi, J.~Mann, and S.~Ott.
\newblock Backwards and forwards relative dispersion in turbulent flow: {An}
  experimental investigation.
\newblock {\em Phys. Rev. E}, 74:016304, 2006.

\bibitem{BIC:2014}
A.~D. Bragg, P.~J. Ireland, and L.~R. Collins.
\newblock Forward and backward in time dispersion of fluid and inertial
  particles in isotropic turbulence.
\newblock {\em arXiv}, page 1403.5502, 2014.

\bibitem{JXPB14}
J.~Jucha, H.~Xu, A.~Pumir, and E.~Bodenschatz.
\newblock Time-reversal-symmetry breaking in turbulence.
\newblock {\em Phys. Rev. Lett.}, 113:054501, 2014.

\bibitem{FF13}
G.~Falkovich and A.~Frishman.
\newblock Single flow snapshot reveals the future and the past of pairs of
  particles in turbulence.
\newblock {\em Phys. Rev. Lett.}, 110:214502, 2013.

\bibitem{FF14}
A.~Frishman and G.~Falkovich.
\newblock New type of anomaly in turbulence.
\newblock {\em Phys. Rev. Lett.}, 113:024501, 2014.

\bibitem{CP2001}
P.~Castiglione and A.~Pumir.
\newblock Evolution of triangles in a two-dimensional turbulent flow.
\newblock {\em Phys. Rev. E}, 64:056303, 2001.

\bibitem{CPS99}
M.~Chertkov, A.~Pumir, and B.~I. Shraiman.
\newblock Lagrangian tetrad dynamics and the phenomenology of turbulence.
\newblock {\em Phys. Fluids}, 11:2394--2410, 1999.

\bibitem{PBX13}
A.~Pumir, E.~Bodenschatz, and H.~Xu.
\newblock Tetrahedron deformation and alignment of perceived vorticity and
  strain in a turbulent flow.
\newblock {\em Phys. Fluids}, 25:035101, 2013.

\bibitem{LOBM07}
B.~L\"uthi, S.~Ott, J.~Berg, and J.~Mann.
\newblock Lagrangian multi-particle statistics.
\newblock {\em J. Turbul.}, 8:45, 2007.

\bibitem{Betchov:1956}
R.~Betchov.
\newblock An inequality concerning the production of vorticity in isotropic
  turbulence.
\newblock {\em J. Fluid Mech.}, 1:497--504, 1956.

\bibitem{Siggia:1981a}
E.~D. Siggia.
\newblock Invariants for the one-point vorticity and strain rate correlation
  functions.
\newblock {\em Phys. Fluids}, 24:1934--1936, 1981.

\bibitem{Kerr:1987}
R.~M. Kerr.
\newblock Histograms of helicity and strain in numerical turbulence.
\newblock {\em Phys. Rev. Lett.}, 59:783--786, 1987.

\bibitem{TKD1992}
A.~Tsinober, E.~Kit, and T.~Dracos.
\newblock Experimental investigation of the field of velocity gradients in
  turbulent flows.
\newblock {\em J. Fluid Mech.}, 242:169--192, 1992.

\bibitem{Vieillefosse:1984}
P.~Vieillefosse.
\newblock Internal motion of a small element of fluid in an inviscid flow.
\newblock {\em Physica A}, 125:150--162, 1984.

\bibitem{Cantwell:1992}
B.~J. Cantwell.
\newblock Exact solution of a restricted {E}uler equation for the velocity
  gradient tensor.
\newblock {\em Phys. Fluids A}, 4:782--793, 1992.

\bibitem{Meneveau:2011}
C.~Meneveau.
\newblock {Lagrangian dynamics and models of the velocity gradient tensor in
  turbulent flows}.
\newblock {\em Annu. Rev. Fluid Mech.}, 43:{219--245}, 2011.

\bibitem{Siggia:1981b}
E.~D. Siggia.
\newblock Numerical study of small-scale intermittency in three-dimensional
  turbulence.
\newblock {\em J. Fluid Mech.}, 107:375--406, 1981.

\bibitem{AKKG87}
W.~T. Ashurst, A.~R. Kerstein, R.~M. Kerr, and C.~H. Gibson.
\newblock Alignment of vorticity and scalar gradient with strain rate in
  simulated {Navier-Stokes} turbulence.
\newblock {\em Phys. Fluids}, 30:2343--2353, 1987.

\bibitem{Tsinober:book}
A.~Tsinober.
\newblock {\em An Informal Conceptual Introduction to Turbulence}.
\newblock Springer, Berlin, 2009.

\bibitem{PN12}
A.~Pumir and A.~Naso.
\newblock Insight on turbulent flows from {Lagrangian} tetrads.
\newblock {\em C. R. Physique}, 13:889--898, 2012.

\bibitem{CM2011}
L.~Chevillard and C.~Meneveau.
\newblock Lagrangian time correlations of vorticity alignments in isotropic
  turbulence: observations and model predictions.
\newblock {\em Phys. Fluids}, 23:101704, 2011.

\bibitem{PW11}
A.~Pumir and M.~Wilkinson.
\newblock Orientation statistics of small particles in turbulence.
\newblock {\em New J. Phys.}, 13:093030, 2011.

\bibitem{NOV14}
R.~Ni, N.~T. Ouellette, and G.~A. Voth.
\newblock Alignment of vorticity and rods with {Lagrangian} fluid stretching in
  turbulence.
\newblock {\em J. Fluid Mech.}, 743:R3, 2014.

\bibitem{NKOV2015}
R.~Ni, S.~Kramel, N.~T. Ouellette, and G.~A. Voth.
\newblock Measurements of the coupling between the tumbling of rods and the
  velocity gradient tensor in turbulence.
\newblock {\em J. Fluid Mech.}, 766:202--225, 2015.

\bibitem{ZGB+11}
R.~Zimmermann, Y.~Gasteuil, M.~Bourgoin, R.~Volk, A.~Pumir, and J.-F. Pinton.
\newblock Rotational intermittency and turbulence induced lift experienced by
  large particles in a turbulent flow.
\newblock {\em Phys. Rev. Lett.}, 106:154501, 2011.

\bibitem{KGBB13}
S.~Klein, M.~Gibert, A.~Berut, and E.~Bodenschatz.
\newblock Simultaneous 3d measurement of the translation and rotation of
  finite-size particles and the flow field in a fully developed turbulent water
  flow.
\newblock {\em Meas. Sci. Technol.}, 24:024006, 2013.

\bibitem{BV12}
G.~Bellani and E.~A. Variano.
\newblock Slip velocity of large neutrally buoyant particles in turbulent
  flows.
\newblock {\em New J. Phys.}, 14:125009, 2012.

\bibitem{CM13}
L.~Chevillard and C.~Meneveau.
\newblock Orientation dynamics of small, triaxial-ellipsoidal particles in
  isotropic turbulence.
\newblock {\em J. Fluid Mech.}, 737:571--596, 2013.

\bibitem{Gust+14}
K.~Gustavsson, J.~Einarsson, and B.~Mehlig.
\newblock Tumbling of small axisymmetric particles in random and turbulent
  flows.
\newblock {\em Phys. Rev. Lett.}, 112:014501, 2014.

\bibitem{MPK+14}
G.~G. Marcus, S.~Parsa, S.~Kramel, R.~Ni, and G.~A. Voth.
\newblock Measurements of the solid-body rotation of anisotropic particles in
  3d turbulence.
\newblock {\em New J. Phys.}, 16:102001, 2014.

\bibitem{Devenish:2013}
B.~J. Devenish.
\newblock Geometrical properties of turbulent dispersion.
\newblock {\em Phys. Rev. Lett.}, 110:064504, 2013.

\bibitem{SPY2013}
B.~L. Sawford, S.~B. Pope, and P.~K. Yeung.
\newblock Gaussian {Lagrangian} stochastic models for multi-particle
  dispersion.
\newblock {\em Phys. Fluids}, 25:055101, 2013.

\bibitem{Yeung:2002}
P.~K. Yeung.
\newblock Lagrangian investigations of turbulence.
\newblock {\em Annu. Rev. Fluid Mech.}, 34:115--142, 2002.

\bibitem{BBC+04}
L.~Biferale, G.~Boffetta, A.~Celani, B.~J. Devenish, A.~Lanotte, and F.~Toschi.
\newblock Multifractal statistics of {L}agrangian velocity and acceleration in
  turbulence.
\newblock {\em Phys. Rev. Lett.}, 93:064502, 2004.

\bibitem{XBOB06}
H.~Xu, M.~Bourgoin, N.~T. Ouellette, and E.~Bodenschatz.
\newblock High order {Lagrangian} velocity statistics in turbulence.
\newblock {\em Phys. Rev. Lett.}, 96:024503, 2006.

\bibitem{ZS:2010}
K.~P. Zybin and V.~A. Sirota.
\newblock {Lagrangian} and {Eulerian} velocity structure functions in
  hydrodynamic turbulence.
\newblock {\em Phys. Rev. Lett.}, 104:154501, 2010.

\bibitem{He:2011}
G.-W. He.
\newblock Anomalous scaling for {Lagrangian} velocity structure functions in
  fully developed turbulence.
\newblock {\em Phys. Rev. E}, 83:025301, 2011.

\bibitem{Borgas:1993}
M.~S. Borgas.
\newblock The multifractal {Lagrangian} nature of turbulence.
\newblock {\em Phil. Trans. R. Soc. Lond. A}, 342:379--411, 1993.

\bibitem{CRL+03}
L.~Chevillard, S.~G. Roux, E.~Leveque, N.~Mordant, J.-F. Pinton, and
  A.~Arneodo.
\newblock Lagrangian velocity statistics in turbulent flows: effects of
  dissipation.
\newblock {\em Phys. Rev. Lett.}, 91:214502, 2003.

\bibitem{CCA+12}
L.~Chevillard, B.~Castaing, A.~Arneodo, E.~Leveque, J.-F. Pinton, and S.~G.
  Roux.
\newblock A phenomenological theory of {Eulerian} and {Lagrangian} velocity
  fluctuations in turbulent flows.
\newblock {\em C. R. Physique}, 13:899--928, 2012.

\bibitem{K62}
A.~N. Kolmogorov.
\newblock A refinement of previous hypotheses concerning the local structure of
  turbulence in a viscous incompressible fluid at high {R}eynolds number.
\newblock {\em J. Fluid Mech.}, 13:82--85, 1962.

\bibitem{SY11}
B.~L. Sawford and P.~K. Yeung.
\newblock Kolmogorov similarity scaling for one-particle {Lagrangian}
  statistics.
\newblock {\em Phys. Fluids}, 23:091704, 2011.

\bibitem{XPFB14}
H.~Xu, A.~Pumir, G.~Falkovich, E.~Bodenschatz, M.~Shats, H.~Xia, N.~Francois,
  and G.~Boffetta.
\newblock Flight-crash events in turbulence.
\newblock {\em Proc. Natl. Acad. Sci. USA}, 111(21):7558--7563, 2014.

\bibitem{Mordant:thesis}
N.~Mordant.
\newblock {\em Mesure lagrangienne en turbulence: mise en {\oe}uvre et
  analyse}.
\newblock PhD thesis, Ecole Normale {Sup\'erieure} de Lyon, 2001.

\bibitem{LN2014}
E.~Leveque and A.~Naso.
\newblock Introduction of longitudinal and transverse {Lagrangian} velocity
  increments in homogeneous and isotropic turbulence.
\newblock {\em Europhys.~Lett.}, 108:54004, 2014.

\bibitem{PXB+14}
A.~Pumir, H.~Xu, G.~Boffetta, G.~Falkovich, and E.~Bodenschatz.
\newblock Redistribution of kinetic energy in turbulent flows.
\newblock {\em Phys. Rev. X}, 4:041006, 2014.

\bibitem{Leray:1934}
Jean Leray.
\newblock Sur le mouvement d'un fluide visqueux emplissant l'espace.
\newblock {\em Acta Mathematica}, 63:193--248, 1934.

\bibitem{Fefferman:2006}
C.~L. Fefferman.
\newblock {\em The millennium prize problems}, chapter Existence and smoothness
  of the {Navier-Stokes} equation, pages 57--67.
\newblock Clay Mathematics Institute, Cambridge, MA, 2006.

\bibitem{TVY2001}
A.~Tsinober, P.~Vedula, and P.~K. Yeung.
\newblock Random {Taylor} hypothesis and the behavior of local and convective
  accelerations in isotropic turbulence.
\newblock {\em Phys. Fluids}, 13:1974--1984, 2001.

\bibitem{gulitski:2007b}
G.~Gulitski, M.~Kholmyansky, W.~Kinzelbach, B.~L{\"u}thi, A.~Tsinober, and
  S.~Yorish.
\newblock Velocity and temperature derivatives in high-{Reynolds}-number
  turbulent flows in the atmospheric surface layer. part 2. accelerations and
  related matters.
\newblock {\em J. Fluid Mech.}, 589:83--102, 2007.

\bibitem{voth:2002}
G.~A. Voth, A.~La~Porta, A.~M. Crawford, J.~Alexander, and E.~Bodenschatz.
\newblock Measurement of particle accelerations in fully developed turbulence.
\newblock {\em J. Fluid Mech.}, 469:121--160, 2002.

\bibitem{BBB+06}
J.~Bec, L.~Biferale, G.~Boffetta, A.~Celani, M.~Cencini, A.~Lanotte,
  S.~Musacchio, and F.~Toschi.
\newblock Acceleration statistics of heavy particles in turbulence.
\newblock {\em J. Fluid Mech.}, 550:349--358, 2006.

\bibitem{QBB+07}
N.~M. Qureshi, M.~Bourgoin, C.~Baudet, A.~Cartellier, and Y.~Gagne.
\newblock Turbulent transport of material particles: {An} experimental study of
  finite size effects.
\newblock {\em Phys. Rev. Lett.}, 99:184502, 2007.

\bibitem{XB08}
H.~Xu and E.~Bodenschatz.
\newblock Motion of inertial particles with size larger than kolmogorov scale
  in turbulent flows.
\newblock {\em Physica D}, 237:2095--2100, 2008.

\bibitem{ZFG+13}
R.~Zimmermann, L.~Fiabane, Y.~Gasteuil, R.~Volk, and J.-F. Pinton.
\newblock Characterizing flows with an instrumented particle measuring
  lagrangian accelerations.
\newblock {\em New J. Phys.}, 15:015018, 2013.

\bibitem{li:2008}
Y.~Li, E.~Perlman, M.~Wan, Y.~Yang, C.~Meneveau, R.~Burns, S.~Chen, A.~Szalay,
  and G.~L. Eyink.
\newblock A public turbulence database cluster and applications to study
  {Lagrangian} evolution of velocity increments in turbulence.
\newblock {\em J. Turbul.}, 9:31, 2008.

\bibitem{yu:2012}
H.~Yu, K.~Kanov, E.~Perelman, J.~Graham, E.~Frederix, R.~Burns, A.~Szalay,
  G.~Eyink, and C.~Meneveau.
\newblock Studying {Lagrangian} dynamics of turbulence using on-demand fluid
  particle tracking in a public turbulence database.
\newblock {\em J. Turbul.}, 13:N12, 2012.

\end{thebibliography}

\end{document}